\documentclass[journal]{IEEEtran}

\usepackage{amsmath,amsfonts}
\usepackage{mathrsfs}
\usepackage{array}
\usepackage{textcomp}
\usepackage{stfloats}
\usepackage{url}
\usepackage{graphicx}
\usepackage{color}
\usepackage{comment}
\usepackage{acronym}
\usepackage{subcaption}
\usepackage{empheq}
\usepackage[font=small]{caption}
\usepackage{enumitem}
\usepackage{cite}
\usepackage{lipsum}
\usepackage{gensymb}
\usepackage{dirtytalk}
\usepackage[T1]{fontenc}
\usepackage{oldgerm}

\newcommand{\bs}[1]{\boldsymbol{#1}}
\newcommand{\wqss}{\bs{\omega}_{\rm \scriptscriptstyle QSS}}
\newcommand{\wu}{\bs{\omega}_\upsilon}
\newcommand{\flux}{\bs{\varphi}}
\newcommand{\voltage}{\bs{\upsilon}}

\newcommand{\vorticity}{\mathbf{w}}
\newcommand{\Clarke}{$\alpha\beta\gamma$}

\newcommand{\wo}{\omega_o}

\newcommand{\sinusoidality}{\varsigma}
\newcommand{\Stan}{\mathcal{S}}

\newcommand{\hilbert}{\mathcal{H}}

\acrodef{pll}[PLL]{Phase-Locked Loop}
\acrodef{qss}[QSS]{Quasi Steady-State}
\acrodef{dfig}[DFIG]{Doubly-Fed Induction Generator}
\acrodef{dft}[DFT]{Discrete Fourier Transform}
\acrodef{pmu}[PMU]{Phasor Measurement Unit}
\acrodef{rl}[RL]{Reinforcement Learning}
\acrodef{thd}[THD]{Total Harmonic Distortion}

\DeclareMathAlphabet{\mathcal}{OMS}{cmsy}{m}{n}

\begin{document}

\title{Sinusoidality Index}

\author{Joan Guti{\'e}rrez-Florensa,~\IEEEmembership{Student,~IEEE,}  {\'A}lvaro Ortega,~\IEEEmembership{Member,~IEEE,} Lukas Sigrist,~\IEEEmembership{Member,~IEEE,} and  Federico Milano,~\IEEEmembership{Fellow,~IEEE}
  \thanks{J.~Guti{\'e}rrez-Florensa and F.~Milano are with the School of Elec.~\& Electron.~Eng., University College Dublin, Dublin, D04V1W8, Ireland.  e-mails: joan.gutierrezflorensa1@ucdconnect.ie, federico.milano@ucd.ie, }%
  \thanks{{\'A}.~Ortega and L.~Sigrist are with School of Engineering, Comillas Pontifical University, 28015, Madrid, Spain. e-mails: \{aortega, lsigrist\}@comillas.edu}%
\thanks{This work is supported by Science Foundation Ireland (SFI) by funding
J.~Guti{\'e}rrez-Florensa and F.~Milano under NexSys project, Grant No. 21/SPP/3756.}
\vspace{-7mm}
}

\maketitle

\begin{abstract}
Maintaining sinusoidal or near-sinusoidal operating conditions in electrical systems is essential, as is their accurate assessment.  This letter proposes a novel metric, namely the \textit{sinusoidality index}, which quantifies the instantaneous deviation of the trajectory of an ac voltage vector with respect to a circle under any periodic operating conditions.  This metric differs from conventional Fourier-based estimations by accounting for the trajectory of the waveform rather than its spectral decomposition.  A variety of examples illustrates the properties of the proposed metric and highlights insights that may not be captured by conventional approaches. 
\end{abstract}


\vspace{-2mm}
\section{Introduction}

Operating at sinusoidal conditions is crucial in electric systems to ensure equipment protection, power quality, and overall system reliability, among other operational considerations.  For this reason, international standards define methods to evaluate waveform distortion and impose limits within which the system can operate safely, e.g.,  \cite{IEEE_Std_Lim}.

However, existing standards estimate the waveform distortion based on its decomposition from the Fourier series and define metrics, such as the \ac{thd},  that rely on the summation of the contribution of each harmonic.  This is the common approach utilized for most power quality indexes proposed in the literature \cite{en14206467}, which are widely used in different applications to evaluate and improve power quality \cite{Yousefpoor2012,Barbie2023}.

While conventional Fourier-based metrics are useful in many applications, they do not account for other conditions that might distort the fundamental signal, e.g., unbalanced conditions, and therefore cannot serve as a generalized metric to evaluate the deviation of a signal from an ideal sinusoid.   Moreover, by the use of Fourier series, these metrics: (i) assume that the signal is periodic at the nominal frequency -- $50$ or $60$ Hz, depending on the region--; and (ii) cannot handle well interharmonics and, more in general, transient conditions. 

This letter uses the geometric and fluid dynamics framework introduced in \cite{Milano2025Lagrange} to propose a novel metric, called \textit{sinusoidality index}, to evaluate the deviation of the trajectory of voltage vectors from an ideal circle.  The metric relies on the estimation of the \ac{qss} frequency, a concept introduced in \cite{gutierrezflorensa2025qss} and that represents the fundamental frequency of the measured voltage.  The proposed metric is able to account for any type of distortion, harmonic and interharmonic, as well as of unbalances.  The proposed index can be also applied to analytic signals if they are properly transformed into planar space vectors.

%--------------------------------------------------

\section{Background and Periodical Conditions }

\label{sec:outlines}

This letter builds on the geometric interpretation of the frequency introduced in \cite{Milano2022Geometric}.  Leveraging on this interpretation, the work in \cite{Milano2025Lagrange}, uses the Lagrange derivative to decompose the geometric frequency, $\wu$, of a vector representing the voltage at a node of three-phase circuit, $\voltage$, into different terms with a clear physical meaning borrowed from fluid mechanics:
\begin{align}
    %\small
    \nonumber
    \wu &= \frac{\voltage\times \voltage'}{|\voltage|^2} \\ \nonumber
    &= \frac{\voltage \times \partial_t\voltage}{|\voltage|^{2}} +
    \frac{\voltage \times (\bs{R}\voltage)}{|\voltage|^{2}}-\frac{1}{2}\frac{(\vorticity\cdot\voltage)\voltage}{|\voltage|^2} + 
    \frac{1}{2}(\nabla\times\voltage) \\   \label{eq:lagrange}
    &= \bs{\omega}_t+\bs{\omega}_r-\frac{1}{2}|\bs{\omega}_\tau|\frac{\voltage}{|\voltage|}+\frac{1}{2}\vorticity \, ,
\end{align}
%\normalsize}
%
where $|\bs v|$ denotes the magnitude and $\bs v'$ the time derivative of $\bs v$ respectively; $\bs{\omega}_t$ is related to the non-stationary conditions of the voltage; $\bs{\omega}_r$ is related to the voltage distortion or \textit{shear strain} characterized by matrix $\bs{R}$; $\bs{\omega}_\tau$ is the torsional frequency; and $\vorticity=\nabla \times \voltage$ is the vorticity, and represents the rotation of a rigid body.  Details on how to obtain all terms that appear in \eqref{eq:lagrange} are given in \cite{Milano2025Lagrange}.  In more than three dimensions, the rigid rotation can be obtained with the exterior gradient ($\nabla\wedge\voltage$) and geometric algebra tools.  For simplicity but without lack of generality, this work focuses only on 3D cases.

Under ideal balanced sinusoidal conditions, $|\bs{\omega}_t|=|\bs{\omega}_r|=|\bs{\omega}_\tau|=0$ and the voltage describes a circular trajectory defined by $1/2|\vorticity|$.  In any other condition, the terms $\bs{\omega}_t$, $\bs{\omega}_r$ and $\bs{\omega}_\tau$ deform the circular path.  The decomposition in \eqref{eq:lagrange} allows extracting the fundamental frequency, given by $1/2|\vorticity|$, from the other terms that compose the geometric frequency.

Nevertheless, estimation of $\vorticity$ by means of its definition based on the curl is not possible since it requires the knowledge of the voltage vector, which is considered as a generalized velocity vector field, as $\voltage(\flux, t)$, where $\flux$ represents the vector of magnetic fluxes that serve as generalized positions.  To overcome this issue, recent work addresses the estimation of vorticity based on a Cauchy theorem that states that vorticity can be measured as the average of $\wu$ \cite{gutierrezflorensa2025qss}, as follows:
\begin{equation}
  \label{eq:vorticity}
  \wqss = \frac{1}{2}\vorticity = 
  \frac{1}{T} \oint_{T}^{}{\wu} \, d\tau \, .
\end{equation}
where $\wqss$ is defined as \textit{quasi-steady-state frequency} in \cite{gutierrezflorensa2025qss} and is the sought fundamental frequency of the voltage; and the period $T$ is defined based on $\wu$ as:
\begin{equation}
  {T=\text{inf}\,\left\{t: t>t_0\, , \int_{t_0}^{t+t_0}|\wu| \, d\tau= 2\pi \right\}} \,.
  \label{eq:Tdef}
\end{equation}
If the trajectory of the voltage is a simple closed curve, the trajectory describes a Jordan's curve, and one can define a unique $T$, which is thus the period of the  fundamental frequency.  Applying \eqref{eq:Tdef} along a moving window allows accounting for variations in time of $T$.  This means that the period utilized in the calculation of $\wqss$ is the \textit{actual} period of the voltage rather than the nominal period of the synchronous reference frequency, as it is assumed in current standards.  A detailed discussion on the challenges of estimating $\wqss$ based on \eqref{eq:vorticity} is given in \cite{gutierrezflorensa2025qss}.  In short, this reference shows that $\wqss$ can be well approximated for stationary as well as slow (e.g., electromechanical) transient conditions.  For fast transients, the estimation of $\wqss$ is still possible but it loses the meaning of fundamental frequency.  Note also that $\wqss$ can be calculated analytically based on the Fourier spectrum of the voltage as obtained, for example, using the DFT.  How to calculate $\wqss$ in the case the voltage is represented as a Fourier series is discussed in \cite{Milano2025Lagrange}.  Using this approach, the limitations of the estimation of $\wqss$ are the same as any other metric based on the Fourier analysis.

\begin{figure}[htb!]
  \centering
  \subfloat[]{\includegraphics[scale=0.39]{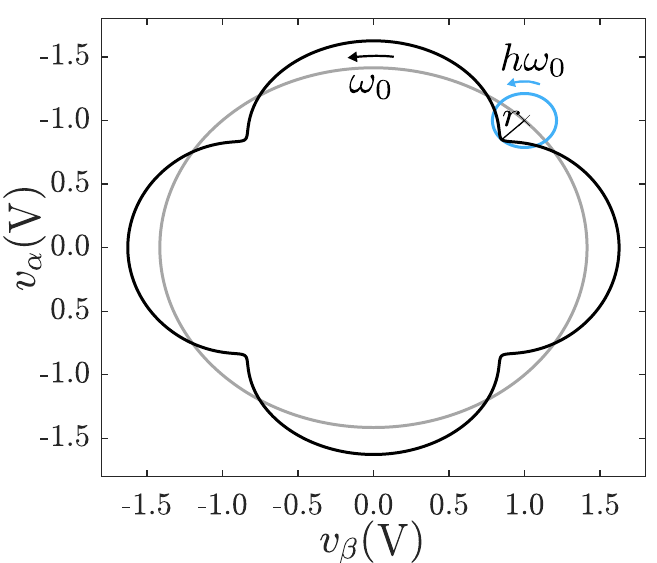}\label{epi_harm}}
  \subfloat[]{\includegraphics[scale=0.39]{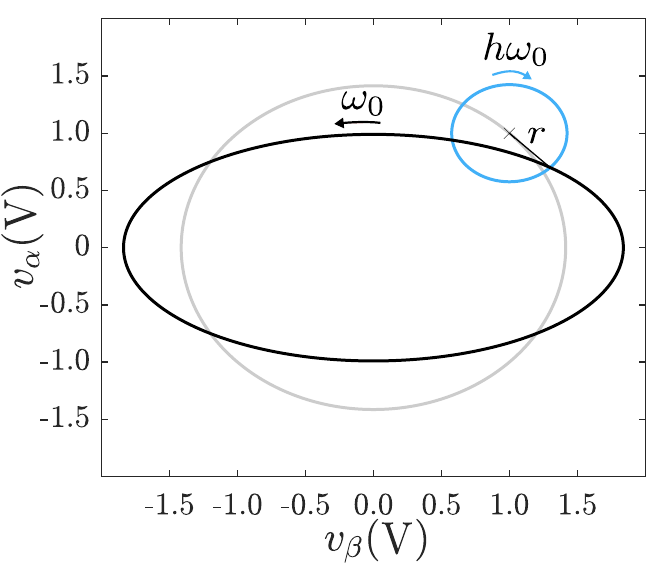}}\label{hypo_unb}
  \centering
  \caption{Comparison of ideal sinusoidal trajectories (gray curves) with actual voltage trajectories (black curves), as a result of epicycles of radius $r=V_h$, where $V_h$ is the magnitude of the harmonic voltage component, for different conditions: (a) a positive-sequence harmonic, $h=5$; (b) an unbalanced sinusoidal  voltage with $h=-1$.}
  \label{fig:trochoids}
\end{figure}

As mentioned above, under periodic conditions, $|\wqss|$ coincides with the fundamental frequency of the voltage as it can be obtained applying the Fourier transform to each element of the voltage vector.  The remaining part of the spectrum of $\voltage$ must then coincide with the other terms of the right-hand side of \eqref{eq:lagrange}.  
As a byproduct, voltage trajectories can be described as a sum of a circular trajectory --- which represents the rigid rotation at the fundamental frequency, $\wo$ --- and a series of \textit{epicycles} given by a series of rotations each at $h\omega_0$ where $h\in\mathbb{Z}$.  Figure \ref{fig:trochoids} illustrates, in the Clarke's $\alpha\beta$-plane, a variety of relevant operating conditions and their link to epicycles.   In this representation, unbalanced conditions
are characterized by an harmonic of order $h=-1$.

In the same vein, single phase voltages can be represented as a three-dimensional vector using the well-known Hilbert transform (see discussion in \cite{MilanoFrenet}), as follows:
\begin{equation}
    \voltage=(\upsilon, \hilbert(\upsilon), 0)^T  \, ,
    \label{eq:v_single_phase_as_vector}
\end{equation}
where $\hilbert(\upsilon)$ is the Hilbert transform of $\upsilon$.  Thus, while in the following we focus exclusively on three-phase systems, single-phase systems can be studied using the same procedure described below.  The only, although relevant, difference is that, due to the nature of the Hilbert transform that is defined in order to remove the negative part of the spectrum of a signal, \eqref{eq:v_single_phase_as_vector} cannot include, by construction, unbalanced conditions.

\color{black}

\section{Sinusoidality Index}

This section presents the proposed sinusoidality index, which quantifies the deviation of a voltage vector from a circular trajectory.  From Fig.~\ref{fig:trochoids}, the intuition behind the proposed metric is straightforward: the proposed sinusoidality index relies on the measurement of the distance between the actual trajectory (black curve) with the ideal circle (gray curve) given by $\frac{1}{2}\vorticity$, estimated with $\wqss$.

The derivation of this metric is build upon the analogy with the dynamical vorticity number, introduced in \cite{truesdell1954kinematics} as a measure of the rotationality of a fluid.  Based on the findings of \cite{Milano2025Lagrange} and \cite{gutierrezflorensa2025qss} we express voltage dynamics, at any moment, as:
\begin{equation}
    \voltage'=\wqss\times\voltage + (\text{distortion terms})\,,
    \label{eq:voltage_prime}
\end{equation}
where $\wqss\times\voltage$ is the ideal rigid motion and the remaining terms account for unbalances, stationary distortions (namely, harmonics) and local time-dependent distortions --- see \cite{Milano2025Lagrange}.

In \cite{truesdell1954kinematics}, the rotationality index, $\Stan_\upsilon$, is defined as:
\begin{equation}
    \Stan_\upsilon=\frac{|\wqss\times\voltage|}{|\voltage'-\wqss\times\voltage|}\,,
\end{equation}
as a scalar that measures the %weight of the dynamics due to the 
contribution of the fundamental frequency on the total voltage dynamics.  To bind and normalize the results, we define the sinusoidality index as:
\begin{equation}
    \sinusoidality_\upsilon=\frac{2}{\pi}\arctan(\Stan_\upsilon)\,.
\end{equation}
Under ideal sinusoidal conditions, the distortion components are null and the voltage dynamics are defined only by $\wqss$, hence, $\sinusoidality_\upsilon=1$. On the other hand, if the signal is irrotational (dc case), $\wqss=0$ and, hence, $\sinusoidality_\upsilon=0$.  Note that $\sinusoidality_\upsilon$ is time-varying under non-ideal conditions (see for instance Section \ref{sec:nonsinus_bal_sys}) and its  average value can be obtained as:
\begin{equation}
    \bar{\sinusoidality}_\upsilon=\frac{1}{T}\oint_T\sinusoidality_\upsilon\,d\tau\,,
    \label{eq: average}
\end{equation}
where $T$ is obtained with \eqref{eq:Tdef}.

\vspace{-3mm}

\section{Analytical and Numerical Case Studies}

\subsection{Three and Single-Phase Sinusoidal Balanced System}

Under sinusoidal balanced conditions we can express voltage in the \Clarke{} frame as:
\begin{equation}
    \voltage=(\upsilon_\alpha,\upsilon_\beta,\upsilon_\gamma)^T=(V\cos\theta,V\sin\theta,0)^T \, .
\end{equation}
This case also illustrates the sinusoidal single-phase case as $\hilbert(V\cos\theta)=V\sin\theta$.  The time derivative of $\voltage$ is:
\begin{equation}
    \voltage'=\omega_0(-V\sin\theta,V\cos\theta,0)^T=\omega_0(-\upsilon_\beta,\upsilon_\alpha,0)^T\,.
\end{equation}

Following the results in \cite{gutierrezflorensa2025qss}, we know that $\wqss=\omega_0(0,0,1)^T$. Hence:
\begin{equation}
    \wqss\times\voltage=\omega_0(-\upsilon_\beta,\upsilon_\alpha,0)^T=\voltage'\,,
\end{equation}
and, consequently, the sinusoidality indices $ \Stan_{\upsilon}$ and $\sinusoidality_\upsilon$ are:
\begin{equation}
    \Stan_{\upsilon} \rightarrow \infty \quad \text{and} \quad \sinusoidality_\upsilon \rightarrow 1\,.
\end{equation}

\vspace{-4mm}

\subsection{Nonsinusoidal Balanced System}
\label{sec:nonsinus_bal_sys}
\begin{figure}[htb!]
  \centering
  \subfloat[]{\includegraphics[scale=0.475]{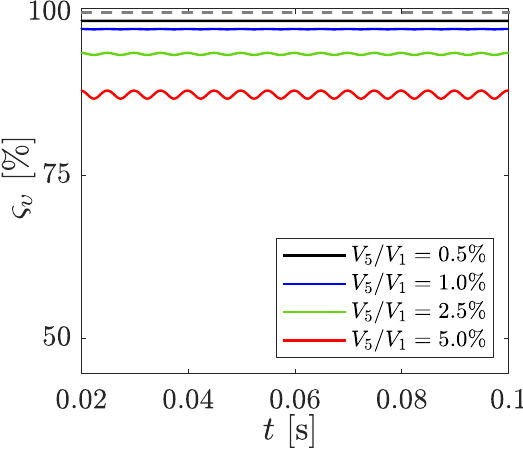}\label{3p_harm_h5}}
    \subfloat[]{\includegraphics[scale=0.475]{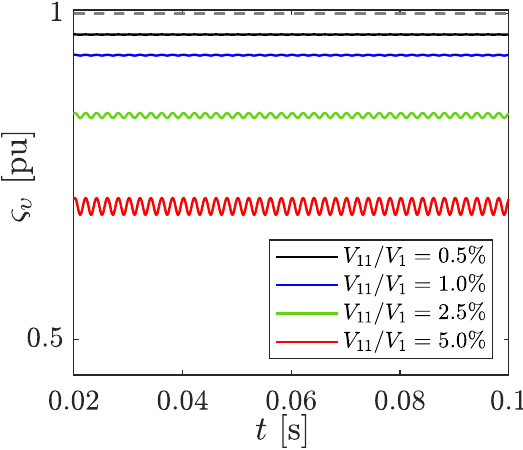}\label{3p_harm_h11}}\\
  \caption{Sinusoidality indexes for a balanced system with one harmonic for different magnitudes of the harmonic: (a) $h=5$, and (b) $h=11$.} 
  \vspace{-3mm}
\end{figure}

In this case we assess the behavior of the proposed metric under the presence of balanced harmonics.  The results for a single harmonic for different magnitudes are shown in Figs.~\ref{3p_harm_h5} and \ref{3p_harm_h11} for $h=5$ and $h=11$ respectively.  Contrary to what is expected from \ac{thd}, higher-order harmonics produce greater distortion from a sinusoid even when the \ac{thd} (in this case equal to $V_h/V$) is the same.
Results discussed below apply equivalently to nonsinusoidal balanced three-phase systems where the voltage can be represented as $\voltage = (\upsilon_{\alpha}, \upsilon_{\beta}, 0)^T$ and to single-phase nonsinusoidal voltage in the form of \eqref{eq:v_single_phase_as_vector}.

As an instantaneous measurement, $\sinusoidality_\upsilon$ is time-varying under non-ideal conditions.  However, the amplitude of the ripple and the average value of $\sinusoidality_\upsilon$ are related as they both depend on $h$ and $V_h$.  For this reason, without loss of information, in the following, we consider the average of the proposed sinusoidality index from \eqref{eq: average}.

\begin{figure}[htb!]
  \centering
  \subfloat[]{\includegraphics[scale=0.475]{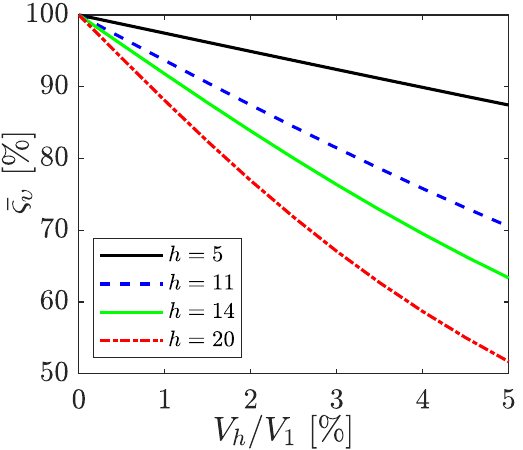}\label{AVG_single_h}}
    \subfloat[]{\includegraphics[scale=0.475]{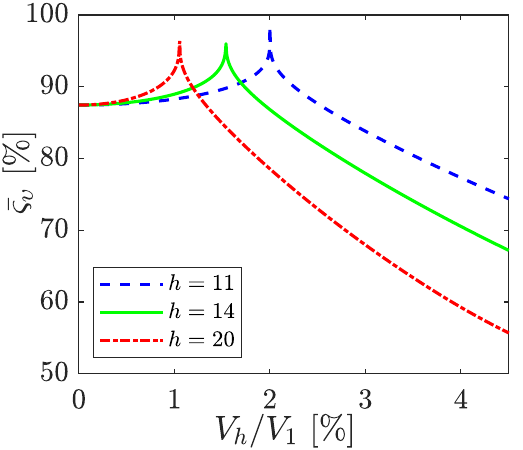}\label{AVG_two_h}}\\
  \caption{Average sinusoidality index of a balanced system with: (a) a single harmonic; and (b) a fixed harmonic $V_5/V_1=5\%$ and an additional harmonic with different $h$ and $V_h$.}
  \label{c}
  \vspace{-3mm}
\end{figure}

Figure \ref{AVG_single_h} shows $\bar{\sinusoidality}_\upsilon$ for different $h$ and $V_h$.  If multiple harmonics are present, results of $\bar{\sinusoidality}_\upsilon$ can be counterintuitive.  Figure \ref{AVG_two_h} shows $\bar{\sinusoidality}_\upsilon$ for a system with two harmonics, one fixed $V_5/V_1=5\%$ and the other one varying in the range shown in the horizontal axis.  According to the \ac{thd}, one would expect the distortion to increase if additional harmonics are included, regardless of its order or magnitude.  However, the behavior of $\bar{\sinusoidality}_\upsilon$ indicates that, in a certain range of magnitudes, the additional harmonic can partially compensate harmonic distortion and reduce the deviation with respect to a circle.

\vspace{-3mm}

\subsection{Three-Phase Sinusoidal Unbalanced }

The sinusoidal unbalanced case is addressed in this section using the symmetric components method.  The results for a variable range of negative sequence amplitudes, denoted with $V_{-1}$; and for fixed zero-sequence amplitudes, denoted with $V_0$, are shown in Fig.~\ref{AVG_unb_sin}.  Results show that, in a certain range of magnitudes, the effect of the negative-sequence can be partially compensated by the zero-sequence component.  This example indicates that $\sinusoidality_\upsilon$ can capture distortions beyond purely harmonic effects, providing a unified framework for waveform distortion estimation under unbalances and harmonics.

\vspace{-3mm}

\subsection{Three-Phase Nonsinusoidal and Unbalanced}

\begin{figure}[htb!]
  \centering
  \subfloat[]{\includegraphics[scale=0.475]{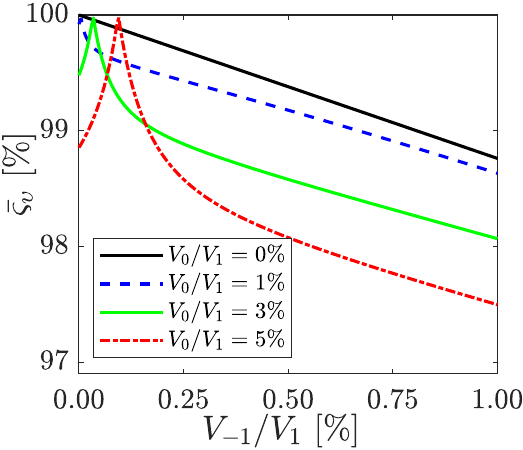}\label{AVG_unb_sin}}
    \subfloat[]{\includegraphics[scale=0.475]{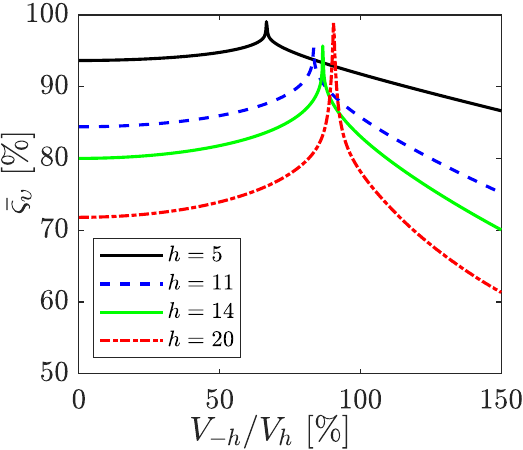}\label{AVG_unb_harm}}\\
  \caption{ Sinusoidality results of a three-phase system with: (a) unbalances;  and (b) a single unbalanced harmonic given by $V_{h\alpha}=V_h+V_{-h}$ and $V_{h\beta}=V_h-V_{-h}$.}
  \vspace{-3mm}
\end{figure}

Finally, we consider the case of a system with unbalanced harmonics.  Figure \ref{AVG_unb_harm} shows the results for $V_h/V_1=2.5\%$ for different $h$ and $V_{-h}$. Results show that below a certain level, the unbalance of the harmonic partially mitigates the effect of the original harmonic distortion.

\vspace{-2mm}

\subsection{Comparison with Conventional Power Quality Indexes}

This last section aims at comparing the proposed index with other power quality indexes commonly used in the literature.
For this comparison, the deviation from the ideal ``circle,'' namely $\sinusoidality=1$, is evaluated.  An analogous approach is used for the crest factor, $C_f$, being the ideal sinusoidal case $C_f=\sqrt{2}$. 

\begin{figure}[htb!]
  \centering
  \subfloat[]{\includegraphics[scale=0.475]{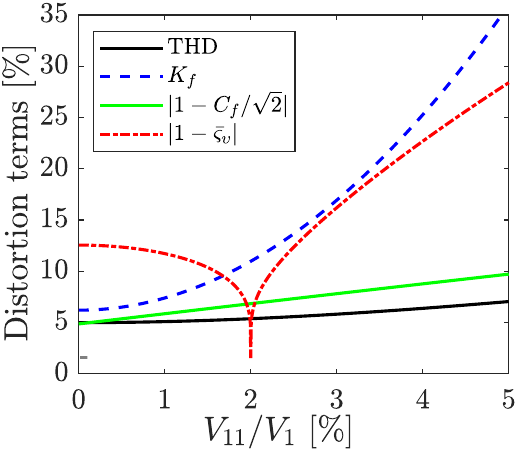}\label{AVG_comparison_h}}
    \subfloat[]{\includegraphics[scale=0.475]{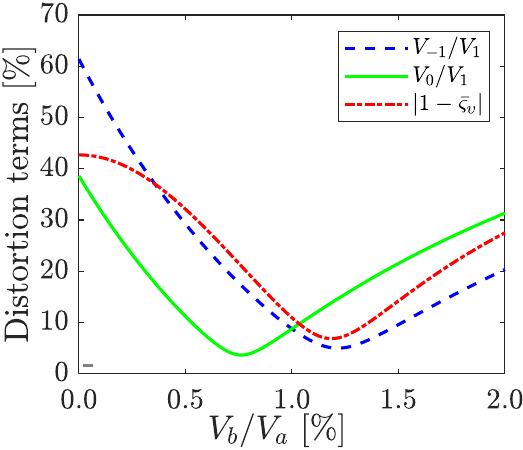}\label{AVG_comparison_unb}}\\
  \caption{Comparison of sinusoidality results with respect to: (a) conventional harmonic indexes under a balanced system with fixed $V_5/V_1$ and variable $V_{11}/V_1$ harmonics; (b) conventional unbalance indexes under a fixed phase unbalance $\phi_c\neq\phi_a=\phi_b$ and a variable magnitude unbalance $V_b\neq V_a =V_c$.}
  %\label{fig:unb_unb_harm}
  \vspace{-3mm}
\end{figure}

Figure \ref{AVG_comparison_h} shows the results for a balanced system with a $5^{\text{th}}$ and an $11^{\text{th}}$ harmonic.  The magnitude of the 5th harmonics is kept constant, whereas the magnitude of the $11^{\text{th}}$ harmonic is varied in the range $V_{11}/V_1 \in [0, 5]$.  The voltage distortion, measured as $|1 - \bar{\sinusoidality}_\upsilon|$, is compared with \ac{thd}, K factor $K_f$, and crest factor $C_f$. 
Results show that only the proposed index is capable to detect the signal quality improvement due to the partial compensation of  $5^{\text{th}}$ harmonic effect through the additional $V_{11}/V_1$ harmonic  at specific range of values. 

Next, we consider an unbalanced voltage with phase shift angles $\phi_a=\phi_b=0$ and $\phi_c=\pi/12$, and magnitudes $V_a=V_c$ with the $V_b$ magnitude changing along the abscissas. This case compares the distortion $|1 - \bar{\sinusoidality}_\upsilon|$ with unbalance indexes, $V_0/V_1$ and $V_{-1}/V_1$, where $V_0$, $V_1$ and $V_{-1}$ represents the magnitudes of the positive, negative and zero sequences of the voltage.  Results are shown in Fig.~\ref{AVG_comparison_unb}.
Differently from conventional indexes, the sinusoidality index takes into account both negative and zero-sequence components.  Results show that, because of the phase unbalance, the minimum distortion is not zero and that the minimum value is not obtained for $V_a = V_b = V_c$.  In fact, the unbalance in the magnitudes of the voltages can partially compensates the distortion due to the phase angle unbalance.  These findings are consistent with the fact that harmonics can partially mutually compensate in a certain range of magnitudes and phase angles.

\color{black}

\vspace{-3mm}

\section{Conclusions}
\label{sec:conclusions}

This letter introduces a novel metric, called \textit{sinusoidality index}, which is a measure of the overall distortion of the trajectory of a voltage vector (or a voltage analytic signal transformed into a plane vector) with respect to a circle.  Leveraging concepts inherited from differential geometry and fluid dynamics, the proposed metric provides a unified measurement of distortion under any periodic condition, even if the period is not perfectly constant.  The proposed approach shows that, under certain conditions, the injection of an additional harmonic may improve the waveform quality.  These cases indicate that conventional metrics, such as the \ac{thd}, do not provide  complete information on the impact of harmonics on the voltage waveform distortion.  The proposed index can be employed for offline power quality analysis -- particularly through its average value -- or formulated as a control objective.  The latter application will be the focus of future work. 

\vspace{-2mm}
% ===================================
%\bibliographystyle{IEEEtran}
%\bibliography{ref}
% Generated by IEEEtran.bst, version: 1.14 (2015/08/26)

% ===================================
\vfill

\newpage

\vfill

\end{document}